# Mimivirus Relatives in the Sargasso Sea

Elodie Ghedin[1]
Jean-Michel Claverie[2]

[1] Department of Parasite and Virus Genomics, The Institute for Genomic Research, 9712 Medical Center Drive, Rockville, MD 20850, USA; Department of Microbiology and Tropical Medicine, George Washington University, Washington DC

[2] Structural and Genomics Information laboratory, CNRS-UPR2589, IBSM, 13402; University of Mediterranee School of Medicine, 13385, Marseille, France

Correspondence to:

Jean-Michel.Claverie@igs.cnrs-mrs.fr

## Summary

The discovery and genome analysis of *Acanthamoeba polyphaga Mimivirus*, the largest known DNA virus, challenged much of the accepted dogma regarding viruses. Its particle size (>400 nm), genome length (1.2 million bp) and huge gene repertoire (911 protein coding genes) all contribute to blur the established boundaries between viruses and the smallest parasitic cellular organisms. Phylogenetic analyses also suggested that the Mimivirus lineage could have emerged prior to the individualization of cellular organisms from the three established domains, triggering a debate that can only be resolved by generating and analyzing more data. The next step is then to seek some evidence that Mimivirus is not the only representative of its kind and determine where to look for new Mimiviridae. An exhaustive similarity search of all Mimivirus predicted proteins against all publicly available sequences identified many of their closest homologues among the Sargasso Sea environmental sequences. Subsequent phylogenetic analyses suggested that unknown large viruses evolutionarily closer to Mimivirus than to any presently characterized species exist in abundance in the Sargasso Sea. Their isolation and genome sequencing could prove invaluable in understanding the origin and diversity of large DNA viruses, and shed some light on the role they eventually played in the emergence of eukaryotes.

## Introduction

The discovery and genome sequence analysis of Mimivirus [1,2], the largest of the Nucleo-cytoplasmic Large DNA Viruses (NCLDV), challenged much of the accepted dogma regarding viruses. Its particle size (>400 nm), genome length (1.2 million bp) and extensive gene repertoire (911 protein coding genes) all contribute to blur the established boundaries between viruses and the smallest parasitic cellular organisms such as Mycoplasma or Nanoarchea [2]. In the universal tree of life, the Mimivirus lineage appears to define a new branch, predating the emergence of all established eukaryotic kingdoms [2]. Although this result is compatible with various hypotheses implicating ancestral DNA viruses in the emergence of eukaryotes [3-5], it requires confirmation from additional data. An urgent task is thus to convince ourselves that Mimivirus is not the sole representative of its kind (i.e. a viral counterpart to the platypus) and to provide some rational guidance as to where to begin the search for eventual new *Mimiviridae*.

Mimivirus was serendipitously discovered within *Acanthamoeba polyphaga*, a free-living ubiquitous amoeba, prevalent in aquatic environments. Phylogenetic analysis of the most conserved genes common to all nucleo-cytoplasmic large double-stranded DNA viruses (NCLDV) [6] positions Mimivirus as an independent lineage, roughly equidistant from the Phycodnaviridae (algal viruses) and Iridoviridae (predominantly fish viruses). Given the ecological affinity of these virus families for the marine environment, we have examined the sequence data set gathered through environmental microbial DNA sampling in the Sargasso Sea [7] to look for possible Mimivirus relatives.

## Results

By comparing Mimivirus ORFs to the Sargasso Sea sequence data set and to all other publicly available sequences, 138 (15%) of the 911 Mimivirus ORFs were found to exhibit their closest match (Blastp E-values ranging from $10^{-74}$ to $10^{-4}$ [8]) to environmental sequences (see Materials and Methods). Even before the discovery of Mimivirus, increasingly complex large double-stranded DNA viruses have been isolated, in particular from unicellular algae. The genome analysis of these Phycodnaviruses revealed a variety of genes encoding enzymes from totally unexpected metabolic pathways [9]. Mimivirus added more unexpected genes (such as translation system components [2]) to this list. As the gene repertoire of these large viruses and the gene content of cellular organisms become increasingly comparable, we have to be cautious in the interpretation of environmental/metagenomics sequence data. To focus our study on environmental organisms most likely to be viruses, we limited further analyses to Mimivirus homologues member of the NCLDV core gene sets [2,6]. These core genes are subdivided into four classes from the most (class I) to least (class IV) evolutionarily conserved [6]. Seven of 10 Mimivirus Class I core genes have their closest homologues in the Sargasso Sea data. This is also the case for 3 of 7 class II core genes, 3 of the 13 class III core genes and 7 of the 16 Class IV core genes (Table 1). Overall, 43% of Mimivirus core genes have their closest homologues in the Sargasso Sea data set. To further assess the viral nature of these unknown microbes, we studied the phylogenetic relationships between the corresponding Mimivirus proteins, their Sargasso Sea homologues, and the closest homologues in other NCLDVs (see Materials and Methods). Figure 1 a-c exhibits three independent phylogenic trees computed using the MEGA3 software [10] for Mimivirus ORFs R449 (unknown function), R429 (unknown function) and L437 (putative virion packaging ATPase). Figure 1a shows that the closest environmental R449 homologues cluster with Mimivirus separately from the known phycodnaviruses, while other Sargasso Sea homologues cluster in a way suggesting the presence of a new clade distinct from Phycodnaviridae. The tree based on R429 and L437 (Fig. 1b, c) similarly suggests the presence of close Mimivirus relatives not belonging to the Phycodnaviridae or Iridoviridae clades.

Another piece of evidence substantiating the existence of an unknown Mimivirus relative in the Sargasso Sea is the discovery of contigs built from the data that contain multiple genes with a high degree of similarity to Mimivirus genes. A spectacular case is illustrated in Figure 2. Here, a 4.5 kb scaffold (See supplementary information) exhibits 4 putative ORFs. When compared to the whole nr database, each of them has as a best match 4 distinct Mimivirus ORFs: thiol oxidoreductase R368 (29% identical, E-value$<10^{-9}$), NTPase-like L377 (25% identical, E-value$<10^{-20}$), unknown function L375 (34% identical, E-value$<10^{-30}$), and DNA repair enzyme L687 (40% identical, E-value$<10^{-62}$). Moreover, the gene order is conserved for three of them (R368, L375, L377). Such colinearity is rarely observed between viral genomes except for members of the same family. Unfortunately, the sequences of these genes are not conserved enough to allow the construction of informative phylogenic trees that would include other NCLDV orthologues.

As of today, genes encoding capsid proteins are among the most unequivocal genes of viral origin. Except for cases of integrated proviral genomes, no cellular homologues of viral capsid proteins have ever been found. During our study, the closest homologues of Mimivirus capsid proteins were found to be capsid protein genes of environmental origin. For example, Mimivirus capsid protein (R441) was found to be 48.5% identical to an unknown environmental sequence, when it is only 36.2% identical to the major capsid protein Vp49 of Chlorella virus CVG-1, its best match among known viruses (Figure 3). As the environmental capsid protein sequence also shares 44.5% identical residues with the CVG-1 Vp49, the corresponding uncharacterized virus appears to lie at an equal evolutionary distance from the *Mimiviridae* and the *Phycodnaviridae*.

## Discussion

Our results predict that DNA viruses of 0.1 to 0.8 microns in size exist in the Sargasso Sea that are evolutionarily closer to Mimivirus than to any presently characterized species. These viruses are abundant enough to have been collected by environmental sampling. It is actually expected that many novel viruses will be encountered in natural waters in which they constitute the most abundant microrganisms [11, 12]. There might be as many as 10 billion virus particles per litre of ocean surface waters [13]. Interestingly, the specialized literature abounds of descriptions of large virus-like particle associated with algae [e.g. 14,15,16], or various marine protists [17,18]. With the exception of Phycodnaviruses [19], the genomic characterization of these viruses has not been attempted. Guided by the results presented here, their isolation and genome sequencing could prove invaluable in understanding the diversity of DNA viruses and the role they eventually played in the evolution of eukaryotes.

## Materials and Methods

The protocols used to collect Sargasso Sea environmental micro-organisms and generate DNA sequences from these samples has been described elsewhere (7). The data analyzed here correspond to "bacteria-sized" organisms that have passed through 3 µm filters and been retained by 0.8 µm to 0.1 µm filters. Mimivirus-like particles (0.8-0.4 µm) belong in this range.

Database similarity searches were performed using the Blast suite of programs (8) (default options) as implemented on the www.giantvirus.org web server and as implemented at The Institute for Genomic Research. Final similarity searches were performed on the non-redundant peptide sequence databases (nr) and environmental data (env-nr) downloaded from the National Institute for Biotechnology Information ftp server (ftp.ncbi.nlm.nih.gov/blast/db/) on March 14, 2005. To avoid missing potential better matches with annotated virus ORFs, all Mimivirus ORFs exhibiting a best match (blosum62 scoring scheme) in env-nr were also searched against all DNA virus genomes using TblastN (peptide query against translated nucleotide sequence). The comprehensive list of Mimivirus ORFs exhibiting a best match in the env-nr database is given in the supplementary material section.

Phylogenetic analyses were conducted using MEGA version 3.0 (10) (option: Neighbor joining, 250 pseudo-replicates, and gaps handled by pairwise deletion). Tree branches were condensed for bootstrap values < 50%.

Only Mimivirus ORFs with best matching homologues in DNA viruses and belonging to the nucleo-cytoplasmic large DNA virus core gene set (2, 6) were analyzed in detail. These ORFs (and matching status) are listed in Table 1 from the most conserved (type 1, in yellow) to the least conserved (type 2 in green, type 3 in blue, and type 4 in white). Phylogenetic analyses were limited to viral homologues and environmental sequences exhibiting a reciprocal best match relationship with the corresponding Mimivirus ORF (putative orthologues) (YES in the rightmost column). The three cases (red lines in Table 1) exhibiting the best bootstrap values are shown in Figure 1. Cases of complex relationships, for instance due to the presence of many paralogues (e.g. capsid proteins), are also indicated. These cases of non-reciprocal best matches are frequent (i.e. the closest homologue of a Mimivirus ORFs being an environmental sequence, but the latter sequence exhibiting a better match with a different ORF in the nr database).

Two environmental sampling contigs - contig IBEA_CTG_1979672 (AACY01022731, GI:44566181) and contig IBEA_CTG_1979673 (AACY01022732, GI:44566179) - are linked in a 4,465 bp scaffold (scaffold IBEA_SCF=2208413) found to contain four ORFs with strong matches to Mimivirus peptides (R368, L377, L375, and L687). The three colinear ORFs (R368, L377, L375) are found on one contig while the orthologue to Mimivirus ORF L687 is found in the second contig. It is conceivable that the lack of colinearity for this fourth ORF is due to an assembly error.

## Acknowledgements

We are indebted to James van Etten for pointing out some ancient observations of very large virus-like particles in algae and marine protists. We thank Stéphane Audic for his help with the www.giantvirus.org server and Hiroyuki Ogata and Vish Nene for reading the manuscript.

Table 1. Matching Status of Mimivirus core genes (type 1 to 4).

| ORF# | Definition | Best score in nr | Best score in DNA viruses | Best score in Sargasso Sea | Status | Reciprocal Best match |
|---|---|---|---|---|---|---|
| L206 | Helicase III / VV D5 | 167- virus | 167 | 214 | Best ENV | YES |
| R322 | DNA pol (B family) extein | 207 | 167 | 238 | Best ENV | YES |
| **L437** | **A32 virion packaging ATPase** | **169 - virus** | **169** | **191** | **Best ENV** | **YES** |
| L396 | VV A18 helicase | 200 -virus | 200 | 187 | - | |
| L425 | Capsid protein | 119 - virus | 117 | 142 | Best ENV | complex |
| R439 | Capsid protein | 164 - virus | 159 | 173 | Best ENV | complex |
| R441 | Capsid protein | 137 - virus | 147 | 209 | Best ENV | complex |
| R596 | E10R-Thiol oxidoreductase | 104 -virus | 105 | 119 | Best ENV | YES |
| R350 | VV D6R - helicase | 170 -virus | 170 | 102 | - | |
| R400 | F10L - prot. Kinase | 86 -virus | 86 | 58 | - | |
| R450 | A1L-transcr factor | 52 -virus | 47 | 65 | Best ENV | |
| R339 | TFII-transcr. factor | 62 | 42 | 66 | Best ENV | |
| L524 | MuT-like NTP PP-hydrolase | 40 | 38 | 39 | - | |
| L323 | Myristoylated virion prot. A | 43 | 42 | 40 | - | |
| R493 | PCNA | 92 | 87 | 154 | Best ENV | YES |
| L312 | Small Ribonucl. reduct | 341 | 338 | 310 | - | |
| R313 | Large Ribonucl. reduct | 766 | 741 | 740 | - | |
| **R429** | **PBCV1-A494R-like** | **152-virus** | **152** | **216** | **Best ENV** | **YES** |
| L37 | BroA, KilA-N | 123-virus | 124 | 65 | - | |
| R382 | mRNA–capping enz. | 86 | 78 | 166 | Best ENV | YES |
| L244 | RNA pol. sub 2 (Rbp2) | 727 | 416 | 508 | - | |
| R501 | RNA pol. sub.1 (Rpb1) | 805 | 415 | 520 | - | |
| R195 | ESV128- Glutaredoxin | 50 | 39 | 49 | - | |
| R622 | S/Y phosphatase | 75 | 73 | 65 | - | |
| R311 | CIV193R BIR domain | 68 | 44 | 51 | - | |
| L65 | Virion memb. prot | 44 | 44 | - | - | |
| R480 | Topoisomerase II | 902 | 717 | 367 | - | |
| L221 | Topoisomerase I bacterial | 528 | 35 | 516 | - | |
| R194 | Topoisomerase I pox-like | 188 | 100 | 145 | - | |
| L364 | SW1/SNF2 helicase | 70-virus | 70 | 72 | Best ENV | YES |
| L4 + 7 par | N1R/P28 DNA binding prot | 123-virus | 124 | 72 | - | |
| L540 | Pre-mRNA helicase - splicing | 256 | 136 | 214 | - | |
| L235 | RNA pol subunit5 | 69 | 38 | 50 | - | |
| R354 | Lambda-type exonuclease | 69-virus | 69 | 154 | Best ENV | YES |
| R343 | RNAse III | 129 | 112 | 131 | Best ENV | YES |
| R141 | GDP mannose 4,6-dehydratase | 294 | 68 | 252 | - | |
| L258 | Thymidine kinase | 151 | 140 | 124 | - | |
| L271 + par | Ankyrin repeats (66 paralogs) | 179 | 152 | 192 | Best ENV | complex |
| R325 | Metal-dependent hydrolase | 69-virus | 69 | 105 | Best ENV | YES |
| L477 | Cathepsin B | 226 | 43 | 47 | - | |
| R497 | Thymidylate synthase | 278 | 242 | 217 | - | |
| **R449** | **Uncharacterized prot.** | **69-virus** | **69** | **129** | **Best ENV** | **YES** |
| R303 | NAD-dependent DNA ligase | 270-virus | 270 | 228 | - | |
| L805 | MACRO domain | 36 | 33 | - | - | |
| R571 L446 | Patatin-like phospholipase | 105 | 80 | 122 | Best ENV | YES |
| R301 | Uncharacterized prot. | 48-virus | 48 | 65 | Best ENV | YES |

# Figure legend

**Figure 1. Phylogenetic evidence of uncharacterized Mimivirus relatives. (a)** Neighbor-joining (NJ) clustering (see Materials and Methods) of Mimivirus R449 ORF with its best matching (≈35% identical residues) environmental homologues (noted Sargasso1 to Sargasso6 according to their decreasing similarity) and closest viral orthologues (28% identical). **(b)** NJ clustering of Mimivirus R429 ORF with its best matching (≈50% identical) environmental homologues (noted Sargasso1 to Sargasso5) and closest viral orthologues (35% identical). **(c)** NJ clustering of Mimivirus putative virion packaging ATPase L437 with its best matching (≈45% identity) environmental homologues (Sargasso1 and Sargasso2) and closest viral orthologues (34% identical). Abbreviations: Phyco: Phycodnavirus; PBCV: *Paramecium bursaria* chlorella virus 1; EsV: *Ectocarpus siliculosus* virus; FsV: Feldmannia sp. virus; HaV: Heterosigma akashiwo virus; Irido: Iridovirus; LCDV: Lymphocystis disease virus 1; Frog: Frog virus 3; Amby: *Ambystoma tigrinum stebbensi* virus; Rana: *Rana tigrina ranavirus*; Chilo: Chilo iridescent virus. Bootstrap values larger than 50% are shown. Branches with lower values were condensed.

**Figure 2. Organization of four Mimivirus ORF best matching homologues in a 4.5 kb environmental sequence scaffold** (approximately to scale). The three colinear Mimivirus homologues are in green. Unmatched ORF extremities are indicated by dots. The two diagonal lines indicate where the two contigs are joined on the scaffold.

**Figure 3. Partial 3-way alignment (N-terminus region) of Mimivirus capsid protein (R441) with it best matching homologues in the NR and Environmental sequence databases.** The Mimivirus R441 protein shares 83/229 (36.2%) identical residues (colored in red or blue) with the major capsid protein Vp49 of Chlorella virus CVG-1 and 111/229 (48.5%) identical residues (indicated in red or green) with the N-terminus of a capsid protein from an unknown large virus sampled from the Sargasso Sea (Accession: EAD00518). On the other hand, the CVG-1 Vp49 and the Sargasso Sea sequence share 44.5 % identical residues. By comparison, the CVG-1 Vp49 protein share 72% of identical residue with PBCV-1 Vp54, its best matching homologue among known phycodnaviruses.

**Figure 1**

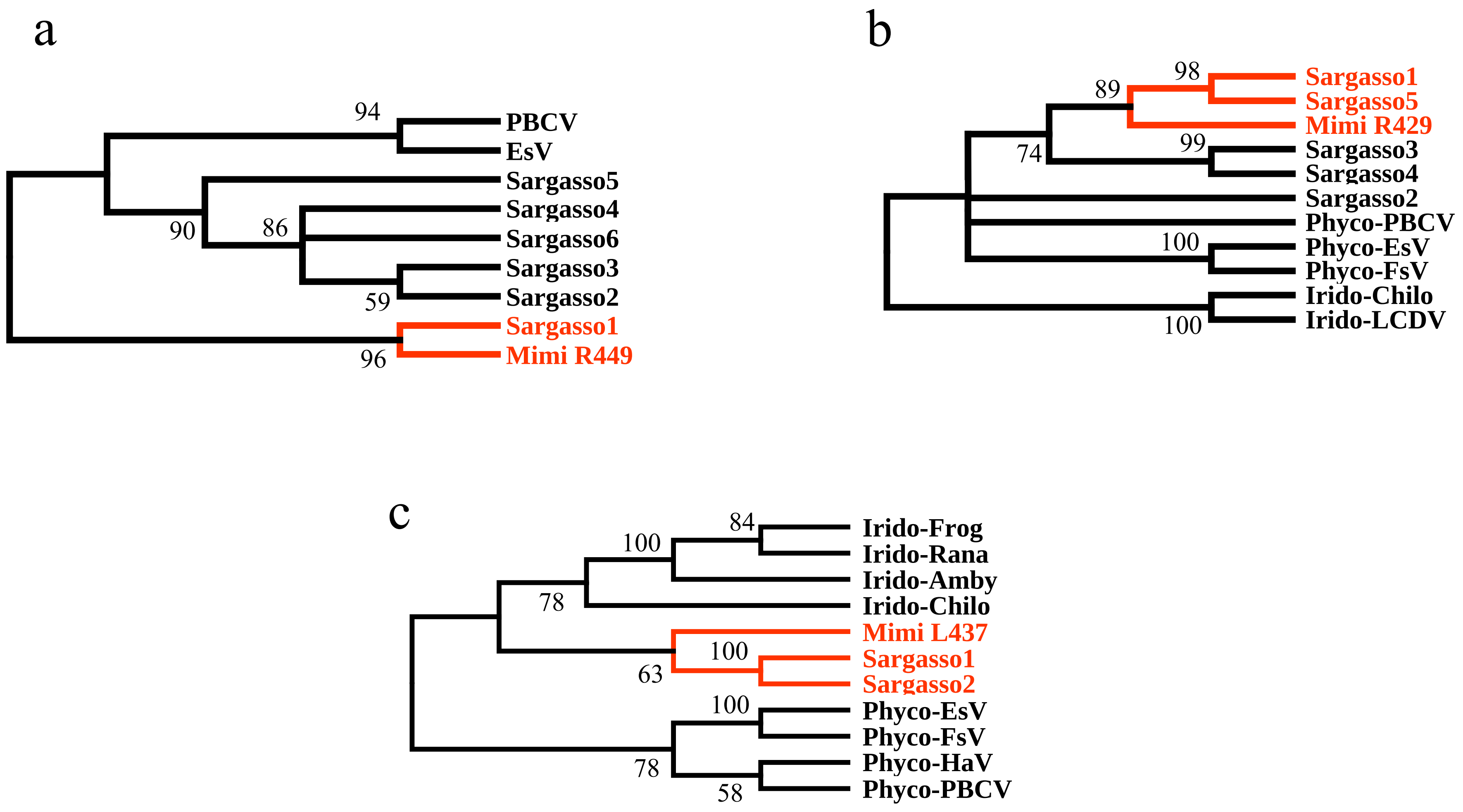

**Figure 2**

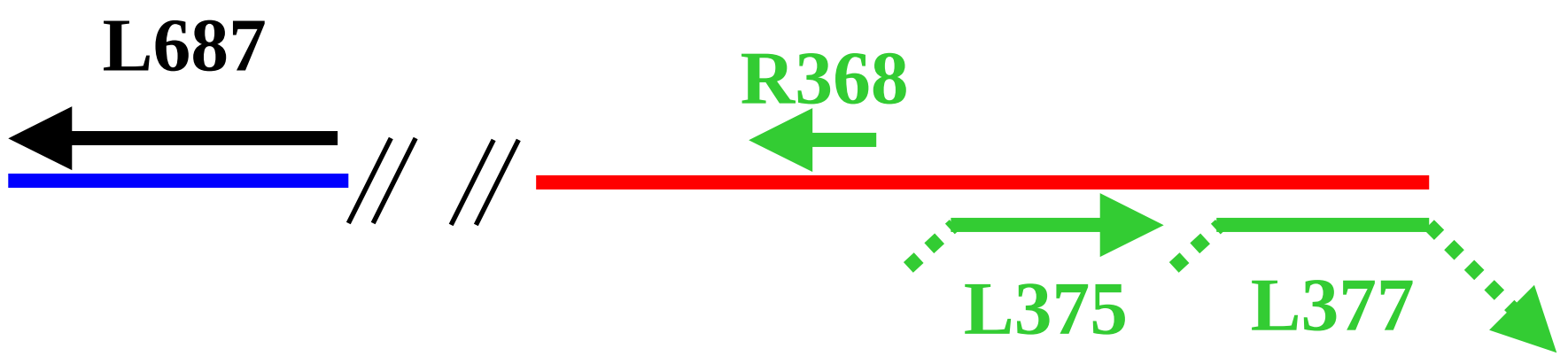

Figure 3

```
CVG1-vp49     MAGGLSQLVAYGAQDVYLTGNPQITFFKTVYRRYTNFAVESIQQTINGSV
MIMI-R441     MAGGIIQLVAYGIQDLYLTGDPQITFFKVVYRRHTNFSVESIIQNFTSVP
Sargasso1     MGGGLMQLVAYGAQDIYLTGNPQITFFKVVYRRHTNFSVESIKQTFNGTA

CVG1-vp49     GFGNKVSTQISRNGDLITDIVVEFVLTKQGPTFY----------------
MIMI-R441     DFGSTVSCTLSKSGDMINKIYVYIELPSVNVFYDESG------NLDKFKK
Sargasso1     DFGKKVSCTISRNGDLVHRIFLQTTLPAQKYDYASAGGGTVTYNSNSNMK

CVG1-vp49     ---------CAEQLLQDVELEIGGQRIDKHYADWFRMYDSLFRMD-----
MIMI-R441     ---FAWVRNIGYALIKDVSIEIGGKLIDKQYGEWMYIWSQVTNKS--DEG
Sargasso1     DGILRWINWVGEKLINYAEIEIGGQRIDKHYGEWLHIWGQLTNTASHDEG

CVG1-vp49     --NDRQNYRRM-----TDFVNDEPATAVKRFYVPLIFFFNQTPGLALPLI
MIMI-R441     LDKMIGNIPLL-----YDFSNGKP---KYSLYVPLEFWFCRNSGLSLPLV
Sargasso1     YQRMVGNIPALTTNVSTNTVAGAAEIKAQDLYVPLQFWFCRNPGLALPLI

CVG1-vp49     ALQYHEVKLYFTLAST--------------VNGITAVEGGAAVTAVAP
MIMI-R441     ALSSSEVKITISFRSAEECYRIGPTHSIEIMEDIVPFEFGDYIEQKIG
Sargasso1     ALQYHEVKINIEFEEL--------------KNLFIAQEKTTAATAVTN
```

Supplementary material

## List of Mimivirus ORFs exhibiting a best match in the env-nr database

| Mimi ORF | Best Env-nr Match ID | Score | E-value | Best nr Match ID | Score | E-value |
|---|---|---|---|---|---|---|
| L18 | gi\|43710841\|gb\|EAF10508.1\| | 137.1 | 4.7e-31 | gi\|51244078\|ref\|YP_063962.1\| | 106.7 | 2.7e-21 |
| L25 | gi\|42923128\|gb\|EAB28610.1\| | 169.1 | 5e-41 | gi\|34876677\|ref\|XP_214012.2\| | 163.3 | 1.1e-38 |
| L75 | gi\|43124292\|gb\|EAC28017.1\| | 51.6 | 1.1e-05 | gi\|13794513\|gb\|AAK39888.1\| | 51.22 | 5.5e-05 |
| L93 | gi\|42923128\|gb\|EAB28610.1\| | 177.9 | 1.6e-43 | gi\|40740451\|gb\|EAA59641.1\| | 156.8 | 1.5e-36 |
| L100 | gi\|42923128\|gb\|EAB28610.1\| | 119 | 1.4e-25 | gi\|40740451\|gb\|EAA59641.1\| | 117.9 | 1.2e-24 |
| L102 | gi\|44215848\|gb\|EAH75003.1\| | 46.98 | 0.00017 | gi\|19703853\|ref\|NP_603415.1\| | 41.2 | 0.035 |
| R106 | gi\|43776260\|gb\|EAF42614.1\| | 139 | 8.1e-32 | gi\|37676542\|ref\|NP_936938.1\| | 107.1 | 1.4e-21 |
| L111 | gi\|43296731\|gb\|EAD14168.1\| | 46.21 | 0.0006 | gi\|16805307\|ref\|NP_473335.1\| | 41.2 | 0.077 |
| L113 | gi\|43770287\|gb\|EAF39597.1\| | 50.45 | 8.8e-06 | gi\|23498951\|emb\|CAD51029.1\| | 43.9 | 0.003 |
| R118 | gi\|44281621\|gb\|EAI17665.1\| | 187.6 | 1.6e-46 | gi\|56965351\|ref\|YP_177083.1\| | 43.51 | 0.014 |
| R132 | gi\|42849963\|gb\|EAA92307.1\| | 98.6 | 4.3e-20 | gi\|33864966\|ref\|NP_896525.1\| | 61.62 | 2.1e-08 |
| R135 | gi\|44399015\|gb\|EAI99687.1\| | 99.37 | 1.4e-19 | gi\|48769830\|ref\|ZP_00274174.1\| | 97.06 | 2.7e-18 |
| L136 | gi\|43375980\|gb\|EAD52870.1\| | 211.5 | 9.9e-54 | gi\|56708496\|ref\|YP_170392.1\| | 107.5 | 7.9e-22 |
| R139 | gi\|42931622\|gb\|EAB32840.1\| | 57 | 1.8e-07 | gi\|30409752\|gb\|AAP32727.1\| | 55.45 | 2e-06 |
| L143 | gi\|43788785\|gb\|EAF48931.1\| | 101.7 | 8e-21 | gi\|54031509\|ref\|ZP_00363643.1\| | 53.91 | 7.4e-06 |
| L174 | gi\|43709436\|gb\|EAF09806.1\| | 48.52 | 0.0001 | gi\|13358136\|ref\|NP_078410.1\| | 45.44 | 0.003 |
| L177 | gi\|43660484\|gb\|EAE85189.1\| | 47.75 | 0.00021 | gi\|56526423\|emb\|CAH77752.1\| | 44.67 | 0.007 |
| L193 | gi\|43116621\|gb\|EAC24307.1\| | 138.7 | 1.7e-31 | gi\|59889773\|emb\|CAH19128.1\| | 110.2 | 2.6e-22 |
| L206 | gi\|44511459\|gb\|EAJ77190.1\| | 210.7 | 2e-53 | gi\|13358409\|ref\|NP_078717.1\| | 163.7 | 1.1e-38 |
| L207 | gi\|43582640\|gb\|EAE46140.1\| | 115.2 | 1.9e-24 | gi\|13358409\|ref\|NP_078717.1\| | 101.3 | 1.1e-19 |
| L208 | gi\|43154232\|gb\|EAC42788.1\| | 54.3 | 6.8e-07 | gi\|23509124\|ref\|NP_701792.1\| | 44.67 | 0.002 |
| L215 | gi\|43198765\|gb\|EAC64922.1\| | 45.44 | 9.8e-05 | gi\|30249486\|ref\|NP_841556.1\| | 43.9 | 0.001 |
| R240 | gi\|43054398\|gb\|EAB93597.1\| | 83.57 | 9.3e-15 | | | |
| L250 | gi\|43527692\|gb\|EAE18541.1\| | 133.7 | 3.8e-30 | gi\|17136758\|ref\|NP_476888.1\| | 121.3 | 7.9e-26 |
| L251 | gi\|44049491\|gb\|EAG84927.1\| | 263.5 | 8.5e-69 | gi\|13357908\|ref\|NP_078182.1\| | 251.5 | 1.4e-64 |
| R267 | gi\|42923128\|gb\|EAB28610.1\| | 114 | 2.9e-24 | gi\|40740517\|gb\|EAA59707.1\| | 111.7 | 5.7e-23 |
| L279 | gi\|42977843\|gb\|EAB55869.1\| | 46.21 | 0.00082 | gi\|23508131\|ref\|NP_700801.1\| | 45.05 | 0.007 |
| R296 | gi\|44416870\|gb\|EAJ11899.1\| | 138.7 | 6.8e-32 | gi\|48845024\|ref\|ZP_00299314.1\| | 132.9 | 1.5e-29 |
| R299 | gi\|43651599\|gb\|EAE80654.1\| | 82.42 | 3.4e-15 | gi\|46440930\|gb\|EAL00231.1\| | 80.88 | 3.7e-14 |
| L300 | gi\|44143264\|gb\|EAH36242.1\| | 59.31 | 1.5e-08 | gi\|50292185\|ref\|XP_448525.1\| | 54.3 | 1.7e-06 |
| R301 | gi\|42942338\|gb\|EAB38176.1\| | 64.7 | 1.8e-09 | gi\|15042460\|gb\|AAK82240.1\| | 48.14 | 0.00068 |
| L315 | gi\|42953235\|gb\|EAB43598.1\| | 88.97 | 5.4e-17 | gi\|56964473\|ref\|YP_176204.1\| | 70.86 | 5.9e-11 |
| L318 | gi\|44268792\|gb\|EAI09019.1\| | 214.2 | 1.5e-54 | gi\|33416901\|gb\|AAH55597.1\| | 160.2 | 1e-37 |
| L320 | gi\|43010993\|gb\|EAB72269.1\| | 101.3 | 8.6e-21 | gi\|30691953\|ref\|NP_174343.2\| | 67.01 | 6.9e-10 |
| R322 | gi\|44500797\|gb\|EAJ70042.1\| | 230.7 | 1.1e-58 | gi\|1655695\|emb\|CAA93738.1\| | 160.2 | 7.5e-37 |
| R325 | gi\|42880571\|gb\|EAB07497.1\| | 98.6 | 3.4e-20 | gi\|2738426\|gb\|AAB94453.1\| | 58.15 | 1.8e-07 |
| R339 | gi\|43154902\|gb\|EAC43116.1\| | 60.46 | 7.5e-09 | gi\|17136888\|ref\|NP_476967.1\| | 58.92 | 7.6e-08 |
| R341 | gi\|42929462\|gb\|EAB31721.1\| | 122.1 | 1.6e-26 | gi\|9628175\|ref\|NP_042761.1\| | 72.02 | 7.5e-11 |
| R343 | gi\|43750296\|gb\|EAF29561.1\| | 131.3 | 2.3e-29 | gi\|30021937\|ref\|NP_833568.1\| | 129 | 4.6e-28 |
| R354 | gi\|43821272\|gb\|EAF65062.1\| | 154.1 | 3.4e-36 | gi\|9631735\|ref\|NP_048514.1\| | 68.94 | 5.8e-10 |
| R355 | gi\|43665722\|gb\|EAE87837.1\| | 121.3 | 1.1e-26 | gi\|450711\|emb\|CAA50819.1\| | 52.37 | 2.4e-05 |
| L364 | gi\|42987215\|gb\|EAB60545.1\| | 72.02 | 1.7e-11 | gi\|4049749\|gb\|AAC97709.1\| | 63.54 | 2.4e-08 |
| R366 | gi\|43171483\|gb\|EAC51308.1\| | 255 | 5.5e-66 | gi\|50302815\|ref\|XP_451344.1\| | 144.1 | 5.6e-32 |
| R368 | gi\|43105861\|gb\|EAC18976.1\| | 113.6 | 4.3e-25 | gi\|13177431\|gb\|AAK14575.1\| | 73.94 | 1.1e-12 |
| L371 | gi\|42973284\|gb\|EAB53609.1\| | 113.6 | 7.7e-24 | gi\|40556241\|ref\|NP_955326.1\| | 109.8 | 4.5e-22 |
| L374 | gi\|43006294\|gb\|EAB69978.1\| | 58.15 | 6.6e-08 | gi\|49528793\|emb\|CAG62455.1\| | 37.35 | 0.452 |
| L375 | gi\|44363842\|gb\|EAI74667.1\| | 131.3 | 1.3e-29 | gi\|450699\|emb\|CAA50807.1\| | 72.4 | 2.9e-11 |
| L377 | gi\|43715140\|gb\|EAF12631.1\| | 141.7 | 4.2e-32 | gi\|37722439\|gb\|AAP33184.1\| | 127.9 | 2.6e-27 |
| R378 | gi\|43174352\|gb\|EAC52755.1\| | 53.53 | 1.2e-06 | gi\|56489446\|emb\|CAI03544.1\| | 49.68 | 6.4e-05 |
| R382 | gi\|44465210\|gb\|EAJ45951.1\| | 166 | 2.1e-39 | gi\|6319713\|ref\|NP_009795.1\| | 85.89 | 1.2e-14 |
| R383 | gi\|43262281\|gb\|EAC96836.1\| | 69.32 | 6e-11 | gi\|17137638\|ref\|NP_477413.1\| | 51.6 | 5.1e-05 |
| L388 | gi\|44253396\|gb\|EAH98417.1\| | 77.8 | 1.1e-13 | gi\|56961986\|ref\|YP_173708.1\| | 75.48 | 2.1e-12 |
| L396 | gi\|43615322\|gb\|EAE62377.1\| | 173.3 | 5.1e-42 | gi\|13177345\|gb\|AAK14489.1\| | 167.5 | 1.1e-39 |
| R398 | gi\|43115033\|gb\|EAC23498.1\| | 120.6 | 1.3e-26 | gi\|46226486\|gb\|EAK87480.1\| | 62.39 | 1.7e-08 |
| R409 | gi\|43205815\|gb\|EAC68464.1\| | 52.37 | 2.4e-06 | gi\|37725924\|gb\|AAO38040.1\| | 38.89 | 0.097 |
| R411 | gi\|43523940\|gb\|EAE16653.1\| | 78.57 | 1.8e-13 | gi\|3116125\|emb\|CAA18875.1\| | 77.8 | 1.2e-12 |
| L417 | gi\|44253353\|gb\|EAH98387.1\| | 102.1 | 1.3e-20 | gi\|9632055\|ref\|NP_048844.1\| | 59.31 | 4e-07 |
| R418 | gi\|44628310\|gb\|EAK59119.1\| | 122.9 | 4.8e-28 | gi\|11498373\|ref\|NP_069601.1\| | 121.3 | 5.8e-27 |
| L425 | gi\|44004806\|gb\|EAG60442.1\| | 142.5 | 8e-33 | gi\|4587052\|dbj\|BAA76601.1\| | 119.8 | 2.2e-25 |
| L426 | gi\|43035635\|gb\|EAB84418.1\| | 68.55 | 5e-11 | gi\|46241679\|gb\|AAS83064.1\| | 63.16 | 7.8e-09 |
| R429 | gi\|43593604\|gb\|EAE51661.1\| | 216.5 | 4.1e-55 | gi\|9632061\|ref\|NP_048850.1\| | 152.5 | 2.9e-35 |
| R430 | gi\|42973834\|gb\|EAB53885.1\| | 53.91 | 1.6e-06 | gi\|9632061\|ref\|NP_048850.1\| | 46.21 | 0.001 |
| L432 | gi\|44173673\|gb\|EAH52299.1\| | 151.4 | 5.4e-36 | gi\|48730983\|ref\|ZP_00264729.1\| | 146.4 | 6.6e-34 |
| R435 | gi\|43287470\|gb\|EAD09576.1\| | 66.63 | 4.8e-10 | gi\|23481897\|gb\|EAA18039.1\| | 48.52 | 0.00054 |
| L437 | gi\|44215544\|gb\|EAH74836.1\| | 177.6 | 1.1e-43 | gi\|16151622\|dbj\|BAB69884.1\| | 157.1 | 6.1e-37 |
| R439 | gi\|43111515\|gb\|EAC21807.1\| | 173.7 | 4.6e-42 | gi\|4587052\|dbj\|BAA76601.1\| | 164.5 | 1.1e-38 |
| R440 | gi\|43011410\|gb\|EAB72486.1\| | 90.12 | 2.9e-16 | gi\|23510178\|ref\|NP_702844.1\| | 78.95 | 2.7e-12 |
| R441 | gi\|43269509\|gb\|EAD00518.1\| | 209.9 | 4.8e-53 | gi\|3341805\|gb\|AAC27492.1\| | 137.9 | 9.4e-31 |
| R443 | gi\|44226821\|gb\|EAH81210.1\| | 63.16 | 1e-09 | gi\|56470459\|gb\|EAL48116.1\| | 52.37 | 6e-06 |
| R445 | gi\|43027882\|gb\|EAB80631.1\| | 50.06 | 4.1e-05 | gi\|34397716\|gb\|AAQ66777.1\| | 48.14 | 0.00061 |
| L446 | gi\|42912213\|gb\|EAB23180.1\| | 95.52 | 7.2e-19 | gi\|47569527\|ref\|ZP_00240206.1\| | 84.73 | 5e-15 |
| R447 | gi\|43125924\|gb\|EAC28806.1\| | 60.46 | 2.9e-09 | gi\|9632052\|ref\|NP_048841.1\| | 46.98 | 0.00014 |
| R449 | gi\|43476579\|gb\|EAD93014.1\| | 122.9 | 1.1e-26 | gi\|13177377\|gb\|AAK14521.1\| | 62.39 | 6.9e-08 |
| R450 | gi\|42883300\|gb\|EAB08831.1\| | 48.91 | 8.6e-05 | gi\|38683713\|gb\|AAR26889.1\| | 40.05 | 0.157 |
| L451 | gi\|44145880\|gb\|EAH37686.1\| | 51.22 | 2.2e-05 | gi\|23612730\|ref\|NP_704269.1\| | 45.82 | 0.004 |
| R453 | gi\|43138976\|gb\|EAC35259.1\| | 74.33 | 2e-12 | gi\|23481840\|gb\|EAA17997.1\| | 51.99 | 4.2e-05 |
| L454 | gi\|43184473\|gb\|EAC57805.1\| | 72.79 | 2.7e-11 | gi\|16805082\|ref\|NP_473111.1\| | 70.09 | 7.1e-10 |
| R468 | gi\|43479103\|gb\|EAD94286.1\| | 46.98 | 0.00017 | | | |
| L471 | gi\|44521596\|gb\|EAJ83983.1\| | 48.91 | 0.0001 | gi\|23510142\|ref\|NP_702808.1\| | 44.67 | 0.008 |
| R472 | gi\|43137277\|gb\|EAC34419.1\| | 111.7 | 7.2e-23 | gi\|33414605\|gb\|AAL38220.2\| | 63.54 | 9.3e-08 |
| L479 | gi\|44350833\|gb\|EAI65324.1\| | 82.03 | 5.2e-15 | gi\|52788091\|ref\|YP_093919.1\| | 70.48 | 6e-11 |
| L483 | gi\|42923128\|gb\|EAB28610.1\| | 161 | 1.9e-38 | gi\|42555731\|gb\|EAA78537.1\| | 157.5 | 8.3e-37 |
| L485 | gi\|44073294\|gb\|EAG98118.1\| | 48.52 | 2.8e-05 | gi\|23509456\|ref\|NP_702123.1\| | 41.2 | 0.015 |
| R489 | gi\|43155682\|gb\|EAC43508.1\| | 67.78 | 8e-11 | gi\|9631920\|ref\|NP_048709.1\| | 47.75 | 0.00032 |
| L491 | gi\|44049493\|gb\|EAG84928.1\| | 61.62 | 7.8e-09 | gi\|56473336\|gb\|EAL50770.1\| | 48.52 | 0.00026 |
| R493 | gi\|43155685\|gb\|EAC43510.1\| | 153.3 | 4.7e-36 | gi\|18726\|emb\|CAA39239.1\| | 88.58 | 5.7e-16 |
| L496 | gi\|43924053\|gb\|EAG16876.1\| | 67.4 | 1.5e-10 | gi\|60468406\|gb\|EAL66411.1\| | 51.99 | 2.6e-05 |
| R502 | gi\|44245809\|gb\|EAH92811.1\| | 46.21 | 0.00033 | gi\|23612359\|ref\|NP_703939.1\| | 38.12 | 0.343 |
| L504 | gi\|43757624\|gb\|EAF33258.1\| | 51.22 | 2.1e-05 | gi\|23619172\|ref\|NP_705134.1\| | 40.05 | 0.192 |
| L507 | gi\|44275242\|gb\|EAI13417.1\| | 99.37 | 3.3e-20 | gi\|9632038\|ref\|NP_048827.1\| | 88.97 | 1.7e-16 |
| R508 | gi\|43850909\|gb\|EAF79670.1\| | 70.09 | 3.3e-11 | gi\|46440772\|gb\|EAL00074.1\| | 43.9 | 0.01 |
| R512 | gi\|43084896\|gb\|EAC08655.1\| | 107.8 | 5.3e-23 | gi\|34333239\|gb\|AAQ64394.1\| | 68.55 | 1.3e-10 |
| L539 | gi\|44612196\|gb\|EAK47432.1\| | 60.08 | 2.5e-08 | gi\|48095940\|ref\|XP_394563.1\| | 53.91 | 6.9e-06 |
| R555 | gi\|44074387\|gb\|EAG98725.1\| | 221.1 | 6.6e-56 | gi\|48477311\|ref\|YP_023017.1\| | 129 | 1.4e-27 |
| R568 | gi\|44444139\|gb\|EAJ31047.1\| | 82.03 | 2.2e-14 | gi\|42782567\|ref\|NP_979814.1\| | 78.18 | 1.3e-12 |
| R569 | gi\|43851406\|gb\|EAF79912.1\| | 46.59 | 0.00037 | gi\|14247226\|dbj\|BAB57617.1\| | 42.36 | 0.027 |
| R571 | gi\|42912213\|gb\|EAB23180.1\| | 103.2 | 2.9e-21 | gi\|54303145\|ref\|YP_133138.1\| | 87.43 | 6.5e-16 |
| R592 | gi\|43246588\|gb\|EAC88951.1\| | 70.48 | 8.4e-11 | gi\|19173110\|ref\|NP_597661.1\| | 67.01 | 3.8e-09 |
| L593 | gi\|44438438\|gb\|EAJ27151.1\| | 128.3 | 6e-29 | gi\|54029386\|ref\|ZP_00361528.1\| | 114.4 | 3.4e-24 |
| R595 | gi\|43331658\|gb\|EAD31349.1\| | 97.44 | 1.6e-19 | gi\|42553185\|gb\|EAA76028.1\| | 90.12 | 9.8e-17 |
| R596 | gi\|43382223\|gb\|EAD56003.1\| | 119.8 | 2.9e-26 | gi\|13177431\|gb\|AAK14575.1\| | 104.4 | 5e-21 |
| R604 | gi\|43049485\|gb\|EAB91155.1\| | 47.37 | 0.00028 | gi\|23484531\|gb\|EAA19833.1\| | 44.28 | 0.01 |
| L620 | gi\|42912213\|gb\|EAB23180.1\| | 57.77 | 2e-07 | gi\|19705025\|ref\|NP_602520.1\| | 55.45 | 3.9e-06 |
| R640 | gi\|44283592\|gb\|EAI18988.1\| | 124.8 | 2.1e-27 | gi\|37519905\|ref\|NP_923282.1\| | 112.8 | 3.4e-23 |
| R648 | gi\|44283592\|gb\|EAI18988.1\| | 167.5 | 2e-40 | gi\|37519905\|ref\|NP_923282.1\| | 122.1 | 3.8e-26 |
| R654 | gi\|43291944\|gb\|EAD11806.1\| | 97.44 | 4.3e-19 | gi\|32263328\|gb\|AAP78373.1\| | 84.34 | 1.5e-14 |
| R667 | gi\|43769283\|gb\|EAF39072.1\| | 55.07 | 2.1e-07 | gi\|52841308\|ref\|YP_095107.1\| | 50.83 | 9.6e-06 |
| L687 | gi\|44566182\|gb\|EAK15053.1\| | 244.6 | 9.5e-64 | gi\|46097173\|gb\|EAK82406.1\| | 214.9 | 3.2e-54 |
| R689 | gi\|44068613\|gb\|EAG95539.1\| | 150.6 | 1.2e-35 | gi\|24378003\|gb\|AAN59275.1\| | 144.1 | 4.4e-33 |
| L690 | gi\|44283592\|gb\|EAI18988.1\| | 152.5 | 6e-36 | gi\|37519905\|ref\|NP_923282.1\| | 104.4 | 7.4e-21 |
| R693 | gi\|43163235\|gb\|EAC47212.1\| | 119.4 | 8.9e-27 | gi\|49235486\|ref\|ZP_00329554.1\| | 115.5 | 3.2e-25 |
| L716 | gi\|44016225\|gb\|EAG66729.1\| | 69.32 | 2.4e-11 | gi\|15921096\|ref\|NP_376765.1\| | 66.24 | 7.4e-10 |
| R730 | gi\|44083879\|gb\|EAH03977.1\| | 199.5 | 4.8e-50 | gi\|6466376\|gb\|AAF12958.1\| | 123.2 | 1.7e-26 |
| R753 | gi\|43500977\|gb\|EAE05248.1\| | 85.89 | 1.7e-15 | gi\|46447199\|ref\|YP_008564.1\| | 71.25 | 1.7e-10 |
| R757 | gi\|43500977\|gb\|EAE05248.1\| | 73.17 | 9.5e-12 | gi\|46447096\|ref\|YP_008461.1\| | 65.86 | 6.2e-09 |
| R758 | gi\|44662698\|gb\|EAK80724.1\| | 54.68 | 4.7e-07 | gi\|50755591\|ref\|XP_414809.1\| | 51.22 | 1.9e-05 |
| R760 | gi\|43408903\|gb\|EAD65356.1\| | 97.44 | 3.4e-19 | gi\|20521133\|dbj\|BAA31672.2\| | 96.67 | 2.3e-18 |
| R777 | gi\|42923128\|gb\|EAB28610.1\| | 95.52 | 6.3e-19 | gi\|18676694\|dbj\|BAB84999.1\| | 90.89 | 6.1e-17 |
| L780 | gi\|43372155\|gb\|EAD50923.1\| | 280.4 | 1.3e-74 | gi\|56472969\|gb\|EAL50422.1\| | 237.7 | 3.7e-61 |
| R791 | gi\|43884722\|gb\|EAF96312.1\| | 79.72 | 3.3e-14 | gi\|32419933\|ref\|XP_330410.1\| | 78.95 | 2.2e-13 |
| R802 | gi\|44389434\|gb\|EAI92933.1\| | 59.69 | 2.7e-08 | gi\|25010522\|ref\|NP_734917.1\| | 57.77 | 3.9e-07 |
| R815 | gi\|44467483\|gb\|EAJ47513.1\| | 109.4 | 9.5e-23 | gi\|52841586\|ref\|YP_095385.1\| | 107.8 | 1.1e-21 |
| R832 | gi\|44335875\|gb\|EAI54742.1\| | 123.2 | 4.5e-27 | gi\|13475556\|ref\|NP_107120.1\| | 112.5 | 3.1e-23 |
| R835 | gi\|43881322\|gb\|EAF94645.1\| | 87.43 | 1.8e-16 | gi\|24646968\|ref\|NP_650415.1\| | 84.34 | 5.8e-15 |
| R837 | gi\|44398294\|gb\|EAI99192.1\| | 86.27 | 1e-15 | gi\|1845265\|gb\|AAB47805.1\| | 84.73 | 1.2e-14 |
| R843 | gi\|44554385\|gb\|EAK06651.1\| | 53.53 | 2.9e-06 | | | |
| R844 | gi\|43318044\|gb\|EAD24737.1\| | 83.57 | 1.3e-15 | gi\|53730512\|ref\|ZP_00151259.2\| | 76.64 | 6e-13 |
| R846 | gi\|43713504\|gb\|EAF11838.1\| | 61.23 | 4.2e-09 | gi\|47214900\|emb\|CAG01031.1\| | 60.46 | 2.5e-08 |
| R847 | gi\|44398294\|gb\|EAI99192.1\| | 74.71 | 2.4e-13 | gi\|28373837\|pdb\|1N0R\|A | 71.63 | 5.2e-12 |
| R848 | gi\|42923128\|gb\|EAB28610.1\| | 110.5 | 1.5e-23 | gi\|58699292\|ref\|ZP_00374082.1\| | 98.98 | 1.7e-19 |
| R850 | gi\|43710841\|gb\|EAF10508.1\| | 65.86 | 1.1e-09 | gi\|48834353\|ref\|ZP_00291368.1\| | 63.16 | 3e-08 |
| R852 | gi\|43536538\|gb\|EAE22952.1\| | 115.9 | 2e-25 | gi\|33860982\|ref\|NP_892543.1\| | 99.37 | 7e-20 |
| R853 | gi\|43166050\|gb\|EAC48610.1\| | 108.2 | 1.2e-23 | gi\|33860982\|ref\|NP_892543.1\| | 86.27 | 2.1e-16 |
| R855 | gi\|44585905\|gb\|EAK28804.1\| | 120.6 | 2.4e-27 | gi\|33860982\|ref\|NP_892543.1\| | 101.3 | 6.2e-21 |
| R877 | gi\|43313336\|gb\|EAD22405.1\| | 95.13 | 2.6e-19 | gi\|26990735\|ref\|NP_746160.1\| | 94.74 | 1.2e-18 |
| L893 | gi\|44629372\|gb\|EAK59899.1\| | 62.77 | 1e-09 | gi\|57101548\|ref\|XP_541839.1\| | 59.31 | 2.7e-08 |
| L894 | gi\|44600496\|gb\|EAK39086.1\| | 109.4 | 7.2e-23 | gi\|45684031\|ref\|ZP_00195462.1\| | 93.59 | 1.6e-17 |
| R901 | gi\|42923128\|gb\|EAB28610.1\| | 162.5 | 5.5e-39 | gi\|42555731\|gb\|EAA78537.1\| | 157.1 | 9.1e-37 |